\documentclass[aps,prc,showpacs,twocolumn]{revtex4-1}
\usepackage{graphicx}
\begin{document}

\title{Gibbs--Tolman approach to the curved interface effects in asymmetric
nuclei}
\author{V.M.~Kolomietz}
\author{A.I.~Sanzhur}
\affiliation{Institute for Nuclear Research, 03680 Kiev, Ukraine}

\begin{abstract}
We redefine the surface tension coefficient and the symmetry energy for an
asymmetric nuclear Fermi-liquid drop with a finite diffuse layer.
Considering two-component charged Fermi-liquid drop and following
Gibbs-Tolman concept, we introduce the equimolar radius $R_{e}$ of sharp
surface droplet at which the surface tension is applied and the radius of
tension surface $R_{s}$ (Laplace radius) which provides the minimum of the
surface tension coefficient $\sigma$. We have shown that the nuclear Tolman
length $\xi$ is negative and the modulus of $\xi$ growth quadratically
with asymmetry parameter $X=(N-Z)/(N+Z)$.
\end{abstract}

\maketitle

\section{Introduction}

The nucleus is a two component, charged system with a finite diffuse layer.
This fact specifies a number of various peculiarities of the nuclear surface
and symmetry energies: dependency on the density profile function, non-zero
contribution to the surface symmetry energy, connection to the nuclear
incompressibility, etc. The additional refinements appear due to the quantum
effects arising from the smallness of nucleus. In particular, the curved
interface creates the curvature correction to the surface energy
$E_{\mathcal{S}}$ and the surface part of symmetry energy $E_{\mathrm{sym}}$ of
order $A^{1/3}$ and can play the appreciable role in small nuclei as well as
in neck region of fissionable nuclei.

The presence of the finite diffuse layer in nuclei creates the problem of
the correct definition of the radius and the surface of tension for a small
drop with a diffuse interface. Two different radii have to be introduced in
this case \cite{gibbs,tolm49}: the equimolar radius $R_{e}$, which gives the
actual size of the corresponding sharp-surface droplet, and the radius of
tension $R_{s}$, which derives, in particular, the capillary pressure.
Bellow we will address this problem to the case of two-component nuclear
drop. In general, the presence of the curved interface affects both the bulk
and the surface properties. The curvature correction is usually negligible
in heavy nuclei. However, this correction can be important in some nuclear
processes. For example the yield of fragments at the nuclear
multifragmentation or the probability of clasterization of nuclei from the
freeze-out volume in heavy ion collisions \cite{kosa12}. In both above
mentioned processes, small nuclei necessarily occur and the exponential
dependence of the yield on the surface tension \cite{lali58} should cause a
sensitivity of both processes to the curvature correction. Moreover the
dependency of the curvature interface effects on the isotopic asymmetry of
small fragments can significantly enhance (or suppress) the yields of
neutron rich isotopes.

In the present paper, we analyze of the interface effects in an asymmetric
nuclear Fermi-liquid drop with a finite diffuse layer. We follow the
ideology of the extended Thomas-Fermi approximation (ETFA) with effective
Skyrme-like forces combining the ETFA and the direct variational method with
respect to the nucleon densities, see Ref. \cite{kosa08}. The proton and
neutron densities $\rho _{p}(\mathbf{r})$ and $\rho _{n}(\mathbf{r})$ are
generated by the diffuse-layer profile functions which are eliminated by the
requirement that the energy of the nucleus should be stationary with respect
to variations of these profiles. In order to formulate proper definition for
the drop radius, we use the concept of the dividing surface, originally
introduced by Gibbs \cite{gibbs}. Following the Gibbs method, which is
applied to the case of two component system, we introduce the superficial
(surface) density as the difference (per unit area of dividing surface)
between actual number of particles $A$ and the number of bulk,
$A_{\mathcal{V}}$, and  neutron excess, $A_{-,\mathcal{V}}$, particles which a
drop would contain if the particle densities were uniform.

The plan of the paper is the following. In Sect.~II we discuss the Gibbs's
derivation of equimolar radius in the case of two-component system with
diffuse layer. We then derive in Sect.~III the surface energy and the
surface contribution to symmetry energy. The relation of the leptodermous
$A^{-1/3}$-expansions for finite nuclei to the nuclear matter equation of
state is discussed in Sect.~IV. Our conclusions are given in Sect.~V.

\section{Dividing surface and equimolar radius in asymmetric nuclei}

We consider first the spherical nucleus at zero temperature, having the
mass number $A=N+Z$, the neutron excess $A_{-}=N-Z$ and the asymmetry
parameter $X=A_{-}/A$. The total binding energy of nucleus is $E$. An actual
nucleus has the finite diffuse layer of particle density distribution.
Thereby, the nuclear size is badly specified. In order to formulate proper
definition for the nuclear radius, we will use the concept of dividing
surface of radius $R$, originally introduced by Gibbs \cite{gibbs}.
Following Refs.~\cite{rowi82,gibbs}, we introduce the formal dividing
surface of radius $R$, the corresponding volume $\mathcal{V}=4\pi R^{3}/3$
and the surface area $\mathcal{S}=4\pi R^{2}$. Note that the dividing
surface is arbitrary but it should be located within the nuclear diffuse
layer.

The energy of a nucleus $E$, as well as the mass number $A$ and the neutron
excess $A_{-}$, are spitted into the volume and surface parts,
\begin{equation}
E=E_{\mathcal{V}}+E_{\mathcal{S}}\ +E_{C},  \label{free}
\end{equation}
\begin{equation}
A=A_{\mathcal{V}}+A_{\mathcal{S}}\ ,\ \ \ 
A_{-}=A_{-,\mathcal{V}}+A_{-,\mathcal{S}}.  \label{parts}
\end{equation}
Here the Coulomb energy $E_{C}$ is fixed and does not depend on the dividing
radius $R$. The bulk energy $E_{\mathcal{V}}$ and the surface energies
$E_{\mathcal{S}}$ can be written as \cite{lali58,rowi82}
\begin{equation}
E_{\mathcal{V}}=\left( -P_{\mathcal{V}}+\lambda\varrho_{\mathcal{V}}+
\lambda_{-}\varrho_{-,\mathcal{V}}\right)\mathcal{V}  \label{vol}
\end{equation}
and
\begin{equation}
E_{\mathcal{S}}=\left(\sigma +\lambda\varrho_{\mathcal{S}}+
\lambda_{-}\varrho_{-,\mathcal{S}}\right)\mathcal{S}.  \label{surf}
\end{equation}
Here $P_{\mathcal{V}}$ is the bulk pressure
\begin{equation}
P_{\mathcal{V}}=-\left.\frac{\partial E_{\mathcal{V}}}{\partial\mathcal{V}}
\right\vert_{A_{\mathcal{V}}},  \label{pv}
\end{equation}
$\sigma$ is the surface tension and
$\varrho_{\mathcal{V}}=A_{\mathcal{V}}/\mathcal{V}$ and
$\varrho_{-,\mathcal{V}}=A_{-,\mathcal{V}}/\mathcal{V}$
are, respectively, the total (isoscalar) and the neutron excess (isovector)
volume densities, $\varrho_{\mathcal{S}}=A_{\mathcal{S}}/\mathcal{S}$ and
$\varrho_{-,\mathcal{S}}=A_{-,\mathcal{S}}/\mathcal{S}$ are the
corresponding surface densities. We have used the isoscalar
$\lambda=(\lambda_{n}+\lambda _{p})/2$ and isovector
$\lambda _{-}=(\lambda_{n}-\lambda _{p})/2$ chemical potentials,
where $\lambda _{n}$ and $\lambda_{p}$ are the chemical potentials
of neutron and proton, respectively. The Coulomb energy $E_{C}$ must be excluded
from the chemical potentials $\lambda$ and $\lambda_{-}$ because of
Eqs.~(\ref{free}), (\ref{vol}) and (\ref{surf}). Namely, 
\begin{equation}
\lambda_{n}=\left.\frac{\partial E}{\partial N}\right\vert_{Z},\quad
\lambda_{p}=\left.\frac{\partial E}{\partial Z}\right\vert_{N}-
\lambda_{C},  \label{lambda}
\end{equation}
where
\[
\lambda_{C}=\left.\frac{\partial E_{C}}{\partial Z}\right\vert_{N}. 
\]
Generally, the realistic (experimental) chemical potentials
$\lambda_{\mathrm{tot},n}$ and $\lambda_{\mathrm{tot},p}$ contain the
contributions of the volume, $\lambda_{\mathrm{vol}}$, surface,
$\lambda_{\mathrm{surf}}$, symmetry, $\lambda_{\mathrm{sym}}$, and Coulomb,
$\lambda_{C}$, parts
\[
\lambda_{\mathrm{tot},n}=\left.\frac{\partial E}{{\partial N}}
\right\vert_{Z}=\lambda_{\mathrm{vol}}+\lambda_{\mathrm{surf}}+
\lambda _{\mathrm{sym}}\ ,
\]
\begin{equation}
\lambda_{\mathrm{tot},p}=\left.\frac{\partial E}{{\partial Z}}
\right\vert_{N}=\lambda_{\mathrm{vol}}+\lambda_{\mathrm{surf}}-
\lambda_{\mathrm{sym}}+\lambda_{C}\ ,  \label{lambda2}
\end{equation}
where
\[
\lambda_{\mathrm{sym}}=2b_{\mathrm{sym}}X 
\]
and $b_{\mathrm{sym}}$ is the symmetry energy. The knowledge of the
chemical potentials $\lambda_{\mathrm{tot},n}$ and $\lambda _{\mathrm{tot},p}$
allows us to evaluate the Coulomb shift $\lambda_{C}$. On the
$\beta$-stability line, the following condition should be satisfied
\begin{equation}
\left.\lambda_{\mathrm{tot},n}-\lambda_{\mathrm{tot},p}
\right\vert_{X=X^{\ast}(A)}=0\ ,  \label{stab}
\end{equation}
and Eq. (\ref{lambda2}) provides the relation
\begin{equation}
\lambda_{C}=4b_{\mathrm{sym}}X^{\ast}\ .  \label{lambdac}
\end{equation}
Here $X^{\ast}=X^{\ast}(A)$ indicates the $\beta$-stability line.

Notation $E_{\mathcal{V}}$ stands for the nuclear matter energy of the
uniform densities $\varrho_{\mathcal{V}}$, $\varrho_{-,\mathcal{V}}$
within the volume $\mathcal{V}$. The state of the nuclear matter
inside the specified volume $\mathcal{V}$ is chosen to have the chemical
potentials $\mu$ and $\mu_{-}$ equal to that of the actual droplet. In
more detail, from the equation of state for the nuclear matter one has
chemical potentials $\mu (\rho ,\rho_{-})$ and $\mu_{-}(\rho ,\rho_{-})$
as functions of the isoscalar, $\rho$, and isovector, $\rho_{-}$,
densities. Then, the following conditions should be fulfilled:
\[
\mu (\rho =\varrho_{\mathcal{V}},\ \rho_{-}=
\varrho_{-,\mathcal{V}})=\lambda\ ,
\] 
\begin{equation}
\mu_{-}(\rho =\varrho_{\mathcal{V}},\ \rho_{-}=
\varrho_{-,\mathcal{V}})=\lambda_{-}   \label{matter}
\end{equation}
to derive the specific values of densities $\varrho_{\mathcal{V}}$ and 
$\varrho_{-,\mathcal{V}}$.

The surface part of the energy $E_{\mathcal{S}}$ as well as the surface
particle number $A_{\mathcal{S}}$ and the surface neutron excess
$A_{-,\mathcal{S}}$ are considered as the excess quantities responsible for
``edge'' effects with respect to the corresponding volume quantities.
Using Eqs.~(\ref{free}) -- (\ref{surf}) one obtains 
\begin{equation}
\sigma =\frac{E-\lambda A-\lambda_{-}A_{-}}{\mathcal{S}}+
\frac{P_{\mathcal{V}}\mathcal{V}}{\mathcal{S}}-
\frac{E_{C}}{\mathcal{S}}=
\frac{\Omega -\Omega_{\mathcal{V}}}{\mathcal{S}}\ .\label{sigma0}
\end{equation}
Here the grand potential $\Omega =E-\lambda A-\lambda_{-}A_{-}-E_{C}$ and
its volume part $\Omega_{\mathcal{V}}=-P_{\mathcal{V}}\mathcal{V}=
E_{\mathcal{V}}-\lambda A_{\mathcal{V}}-\lambda_{-}A_{-,\mathcal{V}}$ were
introduced. From Eq.~(\ref{sigma0}) one can see how the value of the surface
tension depends on the choice of the dividing radius $R$, 
\begin{equation}
\sigma\left[ R\right] =\frac{\Omega}{4\pi R^{2}}+
\frac{1}{3}P_{\mathcal{V}}R\ .  \label{sigmaR}
\end{equation}

Taking the derivative from Eq.~(\ref{sigmaR}) with respect to the formal
dividing radius $R$ and using the fact that observables $E$, $\lambda $,
$\lambda_{-}$ and $P$ should not depend on the choice of the dividing
radius, one can rewrite Eq.~(\ref{sigmaR}) as 
\begin{equation}
P_{\mathcal{V}}=2\,\frac{\sigma\left[ R\right]}{R}+
\frac{\partial }{\partial R}\,\sigma\left[ R\right]\ ,  \label{genlap}
\end{equation}
which is the generalized Laplace equation. The formal values of surface
densities $\varrho_{0,\mathcal{S}}$ and $\varrho_{-,\mathcal{S}}$ can be
found from (\ref{parts}) as
\[
\varrho_{\mathcal{S}}[R]=
\frac{A}{4\pi R^{2}}-\frac{1}{3}\varrho_{\mathcal{V}}R\ ,
\]
\begin{equation}
\varrho_{-,\mathcal{S}}[R]=
\frac{A_{-}}{4\pi R^{2}}-\frac{1}{3}\varrho_{-,\mathcal{V}}R\ .\label{surfden}
\end{equation}
In Eqs.~(\ref{sigmaR}) -- (\ref{surfden}) square brackets denote a formal
dependence on the dividing radius $R$ which is still arbitrary and may not
correspond to the actual physical size of the nucleus. To derive the
physical size quantity an additional condition should be imposed on the
location of dividing surface. In general, the surface energy $E_{\mathcal{S}}$
for the arbitrary dividing surface includes the contributions from the
surface tension $\sigma$ and from the binding energy of particles within
the surface layer. The latter contribution can be excluded for the special
choice of dividing (equimolar) radius $R=R_{e}$ which satisfy the condition 
\begin{equation}
\left(\varrho_{\mathcal{S}}\lambda +
\varrho _{-,\mathcal{S}}\lambda_{-}\right)_{R=R_{e}}=0\ .  \label{emolar}
\end{equation}
Here we use the notation $R_{e}$ by the analogy with the equimolar dividing
surface for the case of the one-component liquid \cite{kosa12,rowi82}. For
the dividing radius defined by Eq.~(\ref{emolar}) the surface energy reads 
\begin{equation}
E_{\mathcal{S}}=\sigma_{e}\mathcal{S}_{e}\ ,  \label{efree}
\end{equation}
where $\sigma _{e}\equiv\sigma (R_{e})$ and $\mathcal{S}_{e}=4\pi R_{e}^{2}$.
Using Eqs.~(\ref{surfden}), (\ref{emolar}), the corresponding volume
$\mathcal{V}_{e}=4\pi R_{e}^{3}/3$ is written as 
\begin{equation}
\mathcal{V}_{e}=\frac{\lambda A+\lambda_{-}A_{-}}
{\lambda \varrho _{\mathcal{V}}+\lambda _{-}\varrho _{-,\mathcal{V}}}\ .
\label{evol}
\end{equation}
As seen from Eqs.~(\ref{matter}), (\ref{evol}), the droplet radius $R_{e}$
is determined by the equation of state for the nuclear matter through the
values of the droplet chemical potentials $\lambda$ and $\lambda_{-}$.

\begin{figure}
\includegraphics[width=\columnwidth]{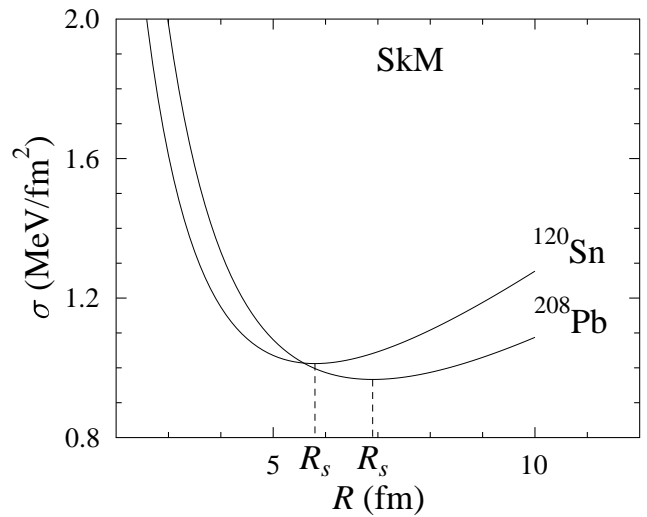}
\caption{
Surface tension $\sigma$ as a function of the dividing
radius $R$ for nuclei $^{120}$Sn and $^{208}$Pb. The calculation was
performed using energy $E$ from Eq.~(\ref{etot1}) and the SkM
force. The Laplace radius $R_{s}$ denotes the dividing radius where
$\sigma$ approaches the minimum value, i.e., the Laplace condition
of Eq. (\ref{p3}) is satisfied.
}
\label{fig1}
\end{figure}

The surface tension $\sigma\left[ R\right]$ depends on the location of the
dividing surface. Function $\sigma\left[ R\right] $ has a minimum at
certain radius $R=R_{s}$ (radius of the surface of tension \cite{rowi82})
which usually does not coincide with the equimolar radius $R_{e}$. The
radius $R_{s}$ (Laplace radius) denotes the location within the interface.
Note that for $R=R_{s}$ the capillary pressure of Eq.~(\ref{genlap})
satisfies the classical Laplace relation 
\begin{equation}
P_{\mathcal{V}}=2\left.\frac{\sigma\left[ R\right]}{R}
\right\vert_{R=R_{s}}\ .  \label{p3}
\end{equation}
The dependence of the surface tension $\sigma\left[ R\right]$ of
Eq.~(\ref{sigmaR}) on the location of the dividing surface for the
nuclei $^{120}$Sn and $^{208}$Pb is shown in Fig.~\ref{fig1}.
%
%

Following Gibbs and Tolman \cite{gibbs,tolm49}, we will assume that the
physical (measurable) value of the surface tension is that taken
at the equimolar dividing surface. We assume, see also Ref.~\cite{rowi82},
that the surface tension $\sigma\equiv\sigma (R_{e})$ approaches the
planar limit $\sigma_{\infty }$ as 
\begin{equation}
\sigma (R_{e})=\sigma_{\infty}\left( 1-\frac{2\xi}{R_{e}}+
\mathcal{O}(R_{e}^{-2})\right)\ ,  \label{sigmaeq}
\end{equation}
where $\xi$ is the Tolman's length \cite{tolm49}. Note that the expression
(\ref{sigmaeq}) can be considered as a particular case of expansion of any
observable $W$ in a finite saturated Fermi-system over the dimensionless
small parameter $r_{0}/R_{e}$, where $r_{0}=(4\pi\rho_{0}/3)^{-1/3}$ and
$\rho_{0}$ is the bulk particle density. Namely,
\begin{equation}
W=W_{\infty}+W_{1}\frac{r_{0}}{R_{e}}+W_{2}
\left(\frac{r_{0}}{R_{e}}\right)^{2}+\ldots\ .  \label{expan}
\end{equation}

Taking Eq.~(\ref{genlap}) for $R=R_{s}$ and comparing with analogous one for 
$R=R_{e}$, one can establish the following important relation
(see Eq.~(\ref{R_e_s}) in Appendix A)
\begin{equation}
\xi =\lim_{A\rightarrow\infty}({R_{e}-R_{s})}\ +\mathcal{O}(X^{2}).
\label{ksi1}
\end{equation}

This result leads to the conclusion that to obtain the non-zero value of
Tolman length $\xi$, and, consequently, the curvature correction
$\Delta\sigma_{\mathrm{curv}}\neq 0$ for a curved surface,
the nucleus must have a finite diffuse surface layer.

\section{Microscopic consideration}

We will perform the numerical calculations using Skyrme type of the
effective nucleon-nucleon interaction. The energy and the chemical
potential for actual droplets can be calculated using a direct variational
method within the extended Thomas-Fermi approximation \cite{kosa08}. The
energy $E$ of the nucleus is given by the following functional 
\[
E=\int d\mathbf{r}\,\left\{ \epsilon_{\mathrm{kin}}[\rho_{n},
\rho_{p};\nabla\rho_{n},\nabla\rho_{p}]+\right.
\]
\begin{equation}
\left.\epsilon_{\mathrm{Sk}}[\rho_{n},\rho_{p};\nabla\rho_{n},
\nabla\rho_{p}]+\epsilon_{C}[\rho_{p}]\right\}\ ,  \label{etot1}
\end{equation}
where $\epsilon_{\mathrm{kin}}[\rho _{n},\rho _{p};\nabla\rho_{n},
\nabla\rho _{p}]$ is the kinetic energy density,
$\epsilon_{\mathrm{Sk}}[\rho_{n},\rho_{p};\nabla\rho_{n},
\nabla\rho_{p}]$ is the potential energy density of Skyrme nucleon-nucleon
interaction and $\epsilon_{C}[\rho _{p}]$ is the Coulomb energy density.
The equilibrium condition can be written as a Lagrange variational problem. Namely, 
\begin{equation}
\delta (E-\lambda_{\mathrm{tot,}n}N-\lambda_{\mathrm{tot,}p}Z)=0\ ,
\label{var1}
\end{equation}%
where the variation with respect to all possible small changes of
$\rho_{n}$ and $\rho_{p}$ is assumed.

Using the trial profile function for the neutron $\rho_{n}(r)$ and proton
$\rho_{p}(r)$ densities and performing the direct variational procedure, we
can evaluate the equilibrium particle densities $\rho (r)=\rho _{n}(r)+
\rho_{p}(r)$ and $\rho_{-}(r)=\rho_{n}(r)-\rho_{p}(r)$, the total energy per
particle $E/A$ and the chemical potentials $\lambda_{\mathrm{tot,}n}$ and
$\lambda_{\mathrm{tot,}p}$ for a fixed asymmetry parameter $X$, see
Ref.~\cite{kosa08} for details. We will also consider the asymmetric nuclear
matter where the energy $E_{\infty}$ is given by 
\begin{equation}
E_{\infty}=\int d\mathbf{r}\,\left\{ \epsilon_{\mathrm{kin}}[\rho_{n},
\rho_{p}]+\epsilon_{\mathrm{Sk}}[\rho_{n},\rho_{p}]\right\}\ .
\label{NM1}
\end{equation}
Here, the kinetic energy density $\epsilon_{\mathrm{kin}}[\rho _{n},\rho_{p}]$
and the potential energy density $\epsilon_{\mathrm{Sk}}[\rho_{n},\rho _{p}]$
do not include the terms which depend on the gradients of nucleon density providing
the bulk particle density $\rho_{0}=\mathrm{const}$. Note also that the Coulomb
energy density $\epsilon _{C}[\rho _{p}]$ does not contribute to the energy
$E_{\infty}$. We will derive the volume (bulk) part of energy $E_{\mathcal{V}}$ as
\begin{equation}
E_{\mathcal{V}}=E_{\infty}\quad \mathrm{and}\quad \varrho_{\mathcal{V}}=\rho _{0}\ .
\label{F1}
\end{equation}
Using the energy $E_{\mathcal{V}}$ from Eq. (\ref{F1}), the above obtained
values of the chemical potentials $\lambda_{n}$ and $\lambda_{p}$ and the
relations
\begin{equation}
\left.\frac{\partial E_{\mathcal{V}}}{\partial A}
\right\vert_{\mathcal{V},A_{-}}=\lambda ,\quad
\left.\frac{\partial E_{\mathcal{V}}}{\partial A_{-}}
\right\vert_{\mathcal{V},A}=\lambda_{-},  \label{basic}
\end{equation}
we will evaluate the equilibrium bulk densities $\varrho _{\mathcal{V}}=\rho_{0}$
and $\varrho_{-,\mathcal{V}}=\rho_{-,0}$.

\begin{figure}
\includegraphics[width=\columnwidth]{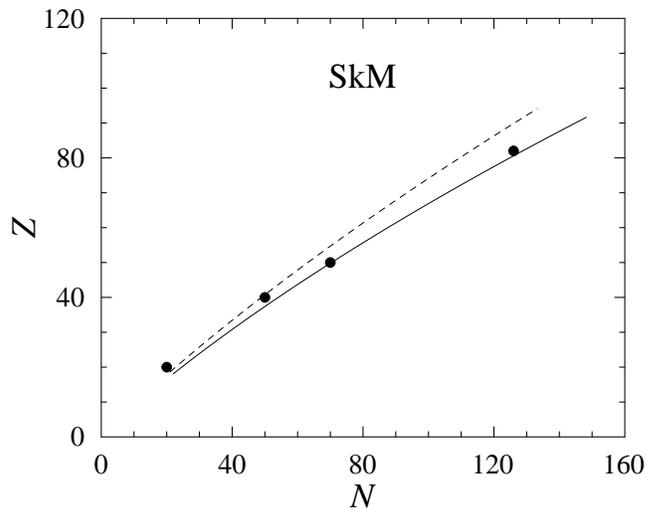}
\caption{
Solid curve is the line of beta stability $Z=Z^{\ast}(N)$ obtained from
Eqs.~(\ref{free}), (\ref{etot1}), (\ref{lambda2}) and (\ref{stab}) for
Skyrme force SkM and dots are the experimental data. The dashed line shows
the ratio of neutrons and protons within the equimolar volume $V_{e}$ of
asymmetric nuclear matter obtained by the Gibbs-Tolman method for nuclei
with $X=X^{\ast}.$
}
\label{fig2}
\end{figure}

The nuclear beta-stability requires the fulfillment of the condition (\ref{stab}).
In Fig.~\ref{fig2} we compare the results for the beta-stability
line $Z=Z^{\ast}(N)$ obtained from Eqs.~(\ref{free}), (\ref{lambda2}) and
(\ref{stab}) with the experimental data (solid dots). One can see that the
solid line gives the acceptable description for the experimental data. Note
that the bulk neutron-proton ratio obtained within the Gibbs-Tolman method
might slightly differ from that of an actual drop. The dashed line in 
Fig.~\ref{fig2} represents function $Z_{\mathcal{V}}(N_{\mathcal{V}})$
which corresponds to $Z^{\ast}(N)$, where the number of protons
$Z_{\mathcal{V}}$ and neutrons $N_{\mathcal{V}}$ are taken for the nuclear
matter within the equimolar volume (\ref{evol}). We can see that for nuclei
along the beta-stability line one has
$X_{\mathcal{V}}=(N_{\mathcal{V}}-Z_{\mathcal{V}})/(N_{\mathcal{V}}+
Z_{\mathcal{V}})<X^{\ast}$. That is because the part of nucleons (mainly neutrons)
are located near the nuclear surface and do not contribute to the volume ratio
$N_{\mathcal{V}}/Z_{\mathcal{V}}$.
%
%

\begin{figure}
\includegraphics[width=\columnwidth]{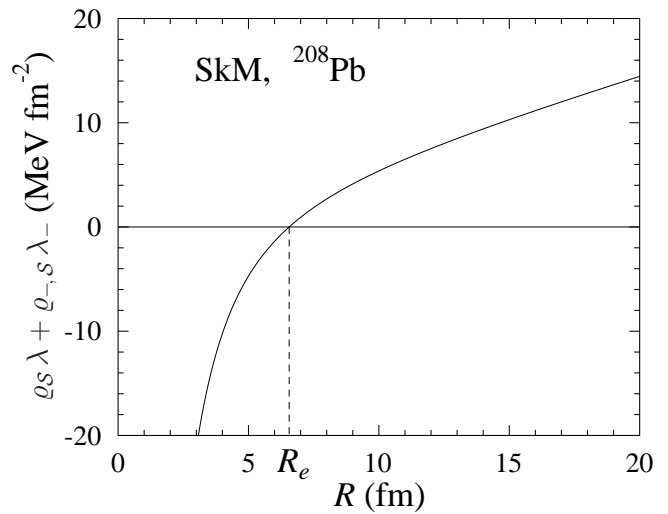}
\caption{
Specific surface particle density $\varrho_{\mathcal{S}}\lambda +
\varrho_{-,\mathcal{S}}\lambda_{-}$ versus dividing
radius $R$ for $^{208}$Pb.
The calculation was performed using the SkM force. $R_{e}$ denotes the
equimolar radius where $\varrho_{\mathcal{S}}\lambda +
\varrho_{-,\mathcal{S}}\lambda_{-}$ becomes zero.
}
\label{fig3}
\end{figure}

\begin{figure}
\includegraphics[width=\columnwidth]{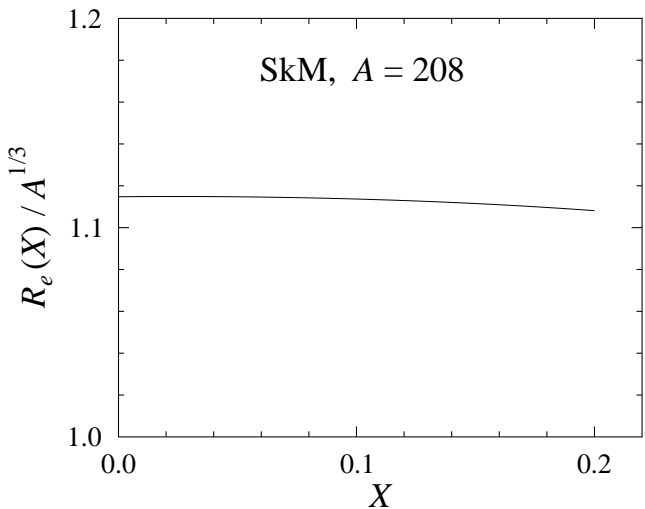}
\caption{
Dependence of the equimolar dividing radius $R_{e}$ on the
asymmetry parameter $X$ for nuclei with $A=208$. The calculation was
performed for Skyrme force SkM.
}
\label{fig4}
\end{figure}

For arbitrary dividing radius $R$ and fixed asymmetry parameter $X$ we
evaluate then the volume, $A_{\mathcal{V}}=4\pi\varrho_{\mathcal{V}}R^{3}/3$
and $A_{-,\mathcal{V}}=4\pi\varrho_{-,\mathcal{V}}R^{3}/3$, the
surface, $A_{\mathcal{S}}=4\pi\varrho_{\mathcal{S}}R^{2}$ and
$A_{-,\mathcal{S}}=4\pi\varrho_{-,\mathcal{S}}R^{2}$, particle numbers and the
volume part of equilibrium energy $E_{\mathcal{V}}$. All evaluated values of 
$E_{\mathcal{V}}[R],$ the bulk densities $\varrho_{\mathcal{V}}$ and
$\varrho_{-,\mathcal{V}}$ and the surface particle densities
$\varrho_{\mathcal{S}}[R]$ and $\varrho_{-,\mathcal{S}}[R]$ depend on the radius $R$
of dividing surface and asymmetry parameter $X$. The actual physical radius
$R_{e}$ of the droplet can be derived by the condition (\ref{emolar}), i.e.,
by the requirements that the contribution to $E_{\mathcal{S}}$ from the bulk
binding energy (term $\sim (\varrho _{\mathcal{S}}\lambda +
\varrho_{-,\mathcal{S}}\lambda _{-})$ in Eq. (\ref{surf})) should be excluded from
the surface energy $E_{\mathcal{S}}$.
In Fig.~\ref{fig3} we represent the calculation of the specific
surface particle density $\varrho_{\mathcal{S}}\lambda +
\varrho_{-,\mathcal{S}}\lambda_{-}$ as a function of the radius $R$ of dividing
surface. Equimolar dividing radius $R_{e}$ in Fig.~\ref{fig3} defines the
physical size of the sharp surface droplet and the surface at which the
surface tension is applied, i.e., the equimolar surface where Eq. (\ref{efree})
is fulfilled.
%
%
The dependence of the equimolar dividing radius $R_{e}$ on the asymmetry
parameter $X$ is shown in Fig.~\ref{fig4}.
%
%

\begin{figure}
\includegraphics[width=\columnwidth]{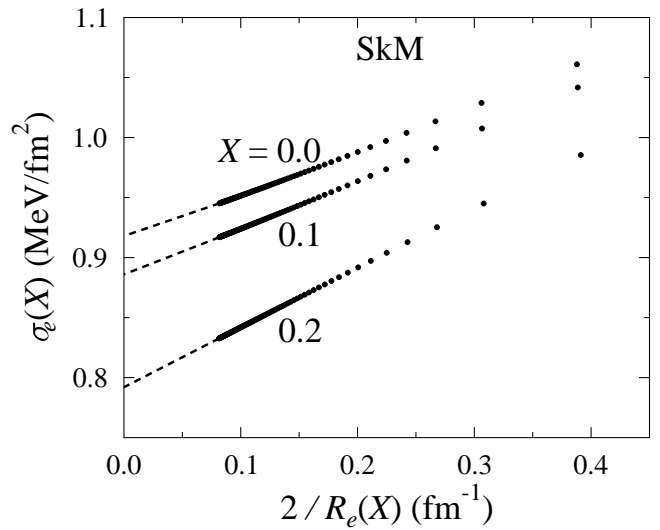}
\caption{
The dependence of the surface tension coefficient $\sigma\left[R_{e},X\right]$
on the equimolar radius $R_{e}$ for different values of the asymmetry parameter
$X$. The calculation was performed for Skyrme force SkM.
}
\label{fig5}
\end{figure}

Note that the value of equimolar radius $R_{e}$, which is derived by Eq.~(\ref{evol}),
is not considerably affected by the Coulomb interaction. We
have also evaluated the values of $R_{e}$ neglecting the Coulomb term in
Eq.~(\ref{etot1}), i.e., assuming $E_{C}=\lambda _{C}=0$. The difference as
compared with data presented in Fig.~\ref{fig4} does not exceed 0.5\%. Omitting the
Coulomb energy contribution to the total energy $E$ of Eq. (\ref{etot1})
and evaluating the bulk energy $E_{\mathcal{V}}$ of Eq. (\ref{F1}), one can
obtain the surface part of energy $E_{\mathcal{S}}=E-E_{\mathcal{V}}$ and
the surface tension coefficient $\sigma\left[ R_{e}\right]$ (\ref{sigma0})
on the equimolar dividing surface for nuclei with different mass number
$A\sim R_{e}^{3}$ and asymmetry parameter $X$. The dependence of the surface
tension coefficient $\sigma\left[ R_{e}\right]$ on the doubled inverse
equimolar radius $2/R_{e}$ (see Eq. (\ref{sigmaeq})) is shown in Fig.~\ref{fig5}.
%
%

\begin{figure}
\includegraphics[width=\columnwidth]{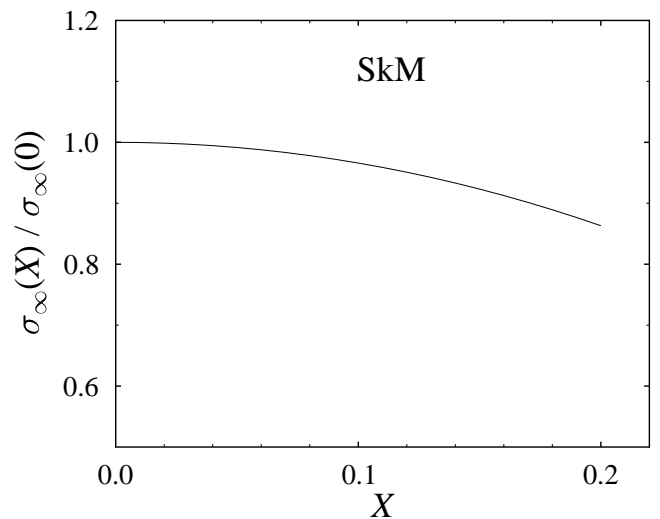}
\caption{
Dependence of the planar surface tension $\sigma_{\infty}(X)$
on the asymmetry parameter $X$. The calculation was
performed for Skyrme force SkM.
}
\label{fig6}
\end{figure}

The surface tension $\sigma\left[ R_{e},X\right] $ approaches the planar
limit $\sigma_{\infty }(X)$ in the limit of zero curvature $2/R_{e}\rightarrow 0$.
As seen from Fig.~\ref{fig5}, the planar limit $\sigma_{\infty}(X)$ depends on the
asymmetry parameter. This dependence
reflects the fact that the symmetry energy $b$ in mass formula contains both
the volume $b_{V}$ and surface $b_{S}$ contributions, see Refs. \cite{dani03,kosa10}
\begin{equation}
b(A)=\ b_{V}+b_{S}\ A^{-1/3}\ .  \label{bsym}
\end{equation}
In Fig.~\ref{fig6} we show the $X$-dependence of the surface tension
$\sigma_{\infty}(X)$. This dependence can be approximated by
\begin{equation}
\sigma_{\infty}(X)=\sigma_{0}+\sigma_{-}X^{2}\ .  \label{sigma1}
\end{equation}%
The dependence of parameters $\sigma_{0}$ and $\sigma_{-}$ on the Skyrme
force parametrization is shown in Table~\ref{tab1}.
%
%

The isovector term $\sigma_{-}$ in the surface tension (\ref{sigma1}) is
related to the surface contribution $b_{S}$ in Eq.~(\ref{bsym}) to the
symmetry energy as
\begin{equation}
b_{S}\approx 4\pi r_{0}^{2}\sigma_{-}\ ,  \label{bsym1}
\end{equation}
see Appendix A, Eq.~(\ref{as1x}). The numerical calculation \cite{kosa10} of the
volume symmetry energy gives for SkM force $b_{V}=$26.5~MeV.
Using Eq.~(\ref{bsym1}), we evaluate the surface-to-volume ratio
$r_{S/V}=|b_{S}/b_{V}| =1.17\div 1.47$ for Skyrme
force parametrizations from Table~\ref{tab1}. Note that in the previous
theoretical calculations, the value of surface-to-volume ratio $r_{S/V}$
varies strongly within the interval $1.6\leq r_{S/V}\leq 2.8$, see
Refs.~\cite{dani03,kosa10,sawy06}.

\begin{figure}
\includegraphics[width=\columnwidth]{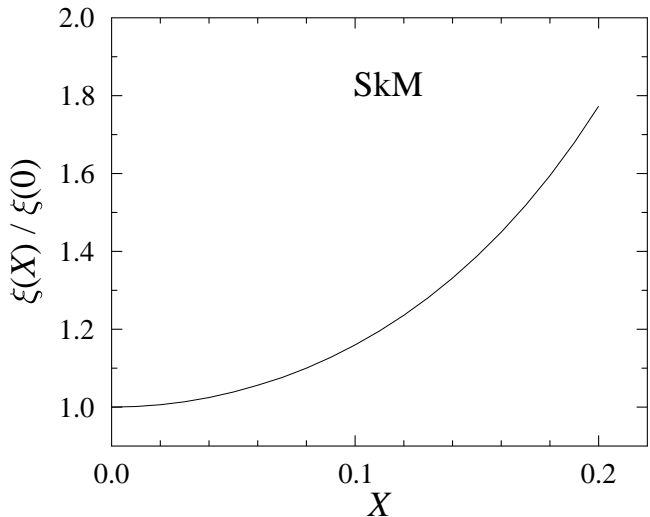}
\caption{
Dependency of the Tolman length $\xi$ on the asymmetry parameter $X$. The
calculation was performed for Skyrme force SkM.
}
\label{fig7}
\end{figure}

The slope of curves $\sigma\left[ R_{e}\right]$ in Fig.~\ref{fig5} gives
the Tolman length $\xi$, see Eq. (\ref{sigmaeq}). The value of the Tolman
length $\xi$ depends significantly on the asymmetry parameter $X$. In 
Fig.~\ref{fig7} we show such kind of dependence obtained from results of 
Fig.~\ref{fig5}.
%
%

As seen from Fig.~\ref{fig7}, one can expect the enhancement of the
curvature effects in neutron rich nuclei. The $X$-dependence of Tolman
length $\xi$ can be approximated as 
\begin{equation}
\xi (X)=\xi_{0}+\xi_{-}X^{2}\ .  \label{xsi1}
\end{equation}%
Both parameters $\xi_{0}$ and $\xi_{-}$ as well as the surface tension
parameter $\sigma_{-}$ are rather sensitive to the Skyrme force
parametrization, see Table~\ref{tab1}.

\section{Nuclear matter equation of state and ($A^{-1/3}$, $X$)-expansions
for finite nuclei}

Bellow we will consider the relation of the nuclear macroscopic
characteristics (surface and symmetry energies, Tolman length,
incompressibility, etc.) to the bulk properties of nuclear matter. Assuming
a small deviations from the equilibrium, the equation of state (EOS) for an
asymmetric nuclear matter can be written in the form expansion around the
saturation point. One has for the energy per particle (at zero
temperature) 
\begin{equation}
\mathcal{E}(\epsilon ,x)=\frac{E_{\infty}}{A}=
\mu_{\infty}+\frac{K_{\infty}}{18}\epsilon^{2}+
b_{\infty}x^{2}+\ldots\ ,  \label{fmatter}
\end{equation}
where
\[
\epsilon =\frac{\rho-\rho_{\infty}}{\rho_{\infty}}\ ,
\  x=\frac{\rho_{-}}{\rho}\ ,
\ \rho =\rho_{n}+\rho_{p}\ ,
\ \rho_{-}=\rho_{n}-\rho_{p}\ ,
\]
$\rho_{\infty}$ is the matter saturation (equilibrium) density,
$\mu_{\infty }$ is the chemical potential, $K_{\infty }$ is the nuclear matter
incompressibility and $b_{\infty}$ is the symmetry energy coefficient (all
values are taken at the saturation point $\epsilon=0$ and $x=0$).
Coefficients of expansion (\ref{fmatter}) are determined through the
derivatives of the energy per particle $\mathcal{E}(\epsilon ,x)$ at the
saturation point: 
\[
\mu_{\infty}=\left.\frac{E_{\infty }}{A}
\right\vert_{\rho =\rho_{\infty},\,x=0}\equiv\mathcal{E}^{(0,0)}\ ,
\]
\begin{equation}
K_{\infty}=9\left.\rho^{2}\frac{\partial^{2}E_{\infty}/A}
{\partial\rho^{2}}
\right\vert_{\rho=\rho_{\infty},\,x=0}\equiv 9\rho_{\infty}^{2}
\mathcal{E}^{(2,0)}\ ,\label{mKJ}
\end{equation}
\begin{equation}
b_{\infty}=\frac{1}{2}\left.\frac{\partial^{2}E_{\infty}/A}
{\partial x^{2}}
\right\vert_{\rho=\rho_{\infty},\,x=0}\equiv\frac{1}{2}
\mathcal{E}^{(0,2)}\ .  \label{b0}
\end{equation}
We use the short notation 
\[
\mathcal{E}^{(n,m)}\equiv\left.\frac{\partial^{n+m}\mathcal{E}}
{\partial\epsilon^{n}\partial x^{m}}\right\vert_{\epsilon=0,\,x=0}\ .
\]
Some coefficients $\mathcal{E}^{(n,m)}$ are vanishing. From the condition of
minimum of $\mathcal{E}(\epsilon ,x)$ at the saturation point one has
$\mathcal{E}^{(1,0)}=\mathcal{E}^{(0,1)}=0$. Odd derivatives with respect to
$x$, i.e., $\mathcal{E}^{(n,m)}$ for odd $m$, also vanish because of the
charge symmetry of nuclear forces.

Using $\mathcal{E}(\epsilon ,x)$, one can also evaluate chemical
potentials $\mu$, $\mu_{-}$ and pressure $P$ of the nuclear matter beyond
the saturation point. Namely,
\[
\mu (\epsilon ,x) = \left.\frac{\partial E_{\infty }}{\partial A}
\right\vert_{A_{-},V}=
\frac{\partial}{\partial\epsilon}(1+\epsilon )\mathcal{E}-
x\frac{\partial\mathcal{E}}{\partial x}\ ,
\]
\begin{equation}
\mu_{-}(\epsilon ,x)=\left.\frac{\partial E_{\infty }}{\partial A_{-}}
\right\vert_{A,V}=\frac{\partial\mathcal{E}}{\partial x}\ ,\label{mu01p} 
\end{equation}
\begin{equation}
P(\epsilon ,x) =-\left.\frac{\partial E_{\infty}}{\partial V}
\right\vert_{A,A_{-}}=\rho_{\infty}(1+\epsilon )^{2}
\frac{\partial\mathcal{E}}{\partial\epsilon}\ .  \label{pex1}
\end{equation}

Similarly to Eq.~(\ref{fmatter}), in a finite uncharged system the energy
per particle $E/A$ (we use $A=N+Z$, $A_{-}=N-Z$, $X=A_{-}/A$) of the
finite droplet is usually presented as ($A^{-1/3}$, $X$)-expansion around
infinite matter using the leptodermous approximation 
\[
E\equiv E(X,A^{-1/3})=a_{V}+X^{2}b_{V}+
\]
\begin{equation}
A^{-1/3}(a_{S}+X^{2}\,b_{S}+a_{c}A^{-1/3}+X^{2}b_{c}A^{-1/3}) \label{eldm1}
\end{equation}
\[
=a_{V}+a_{S}A^{-1/3}+a_{c}A^{-2/3}+
\]
\begin{equation}
X^{2}(b_{V}+b_{S}A^{-1/3}+b_{c}A^{-2/3})
\label{eldm2}
\end{equation}
where $a_{V}$, $a_{S}$ and $a_{c}$ are, respectively, the volume, surface
and curvature energy coefficients, $b_{V}$, $b_{S}$ and $b_{c}$ are,
respectively, the volume, surface and curvature symmetry coefficients. The
nuclear chemical potentials $\lambda$ and $\lambda_{-}$ are derived as
\[
\lambda (X,A^{-1/3})=E/A-\frac{1}{3}\frac{\partial\,E/A}{\partial A^{-1/3}}-
X\frac{\partial \,E/A}{\partial X}\ ,
\]
\begin{equation}
\lambda_{-}(X,A^{-1/3})=\frac{\partial\,E/A}{\partial X}\ .  \label{lambda01}
\end{equation}
Following Gibbs-Tolman method, one can derive the actual nuclear matter
densities $\rho$ and $\rho_{-}$ from the conditions 
\[
\mu (\epsilon ,x)=\lambda (X,A^{-1/3})\ ,
\]
\begin{equation}
\mu_{-}(\epsilon ,x)=\lambda_{-}(X,A^{-1/3})\ .  \label{eosmatt}
\end{equation}
Using Eq.~(\ref{eosmatt}), one can establish the relation of the macroscopic
energy coefficients in the liquid drop model expansion Eq.~(\ref{eldm1}) to
the nuclear matter parameters in EOS (\ref{fmatter}),
see Eqs.~(\ref{av1x}) -- (\ref{R_e_s}) of Appendix A.
The results of numerical calculations
of relevant quantities are represented in Tables~\ref{tab1} and \ref{tab2}.

\noindent
\begin{table}
\caption{
Nuclear bulk parameters for different Skyrme forces.
}
\label{tab1}
\begin{tabular}{lllll}
\hline\hline\noalign{\smallskip}
& ~SkM~~~~ & ~SkM*~~~ & SLy230b & ~~T6 \\
\noalign{\smallskip}\hline\hline\noalign{\smallskip}
$\mu_{\infty}$ (MeV)                & -15.77 & -15.77 & -15.97 & -15.96 \\
$\rho_{\infty}$ (fm$^{-3}$)        & ~0.1603 & ~0.1603 & ~0.1595 & ~0.1609 \\
$K_{\infty}$ (MeV)                 & ~216.6  & ~216.6  & ~229.9  & ~235.9 \\
$K_{3}$ (MeV)                         & ~913.5 & ~913.5 & ~1016. & ~1032. \\
$K_{\mathrm{sym}}$ (MeV)           & -148.8 & -155.9 & -119.7 & -211.5 \\ 
$b_{\infty}$ (MeV)                   & ~30.75 & ~30.03 & ~32.01 & ~29.97 \\
$L_{\infty}$ (MeV)                  & ~49.34 & ~45.78 & ~45.97 & ~30.86 \\
$\sigma_{0}$ (MeV$\cdot$fm$^{-2}$)\ \ \  & ~0.9176 & ~0.9601 & ~1.006 & ~1.021 \\
$\xi_{0}$ (fm)                       & -0.3565 & -0.3703 & -0.3677 & -0.3593 \\
$\sigma_{-}$ (MeV$\cdot$fm$^{-2}$)  & -3.118 & -3.094 & -3.131 & -2.413 \\
$\xi_{-}$ (fm)                      & -5.373 & -5.163 & -4.590 & -2.944 \\
\noalign{\smallskip}\hline
\end{tabular}
\end{table}

\noindent
\begin{table}
\caption{
Mass formula coefficients for finite nuclei.
}
\label{tab2}
\begin{tabular}{lllll}
\hline\hline\noalign{\smallskip}
& ~SkM~~~~ & ~SkM*~~~ & SLy230b & ~~T6 \\
\noalign{\smallskip}\hline\hline\noalign{\smallskip}
$a_{V}$ (MeV) & -15.8 & -15.8 & -16.0 & -16.0 \\
$a_{S}$ (MeV) & ~15.0 & ~15.7 & ~16.5 & ~16.7 \\
$a_{c}$ (MeV) & ~7.30 & ~7.92 & ~8.26 & ~8.16 \\
$b_{V}$ (MeV) & ~30.8 & ~30.0 & ~32.0 & ~30.0 \\
$b_{S}$ (MeV) & -44.2 & -44.1 & -44.9 & -35.1 \\
$b_{c}$ (MeV) & ~35.7 & ~35.1 & ~28.6 & ~17.3 \\
$r_{S/V}=|b_{S}/b_{V}|$\ \  & ~1.44 & ~1.47 & ~1.40 & ~1.17 \\
\noalign{\smallskip}\hline
\end{tabular}
\end{table}

The value of the Tolman length $\xi_{0}$ can be related to the nuclear
matter incompressibility $K_{\infty}$ and the surface tension coefficient
$\sigma$ \cite{blku06}. Let us consider the expansion like (\ref{expan})
around the equilibrium state of the symmetric nuclear matter for the bulk
density and the chemical potential:
\[
\varrho_{\mathcal{V}}=\rho_{\infty}+\rho_{1}\frac{r_{0}}{R_{e}}+
\rho_{2}\left(\frac{r_{0}}{R_{e}}\right)^{2}+\ldots\ ,
\]
\begin{equation}
\lambda =\lambda_{\infty}+\lambda_{1}\frac{r_{0}}{R_{e}}+
\lambda_{2}\left(\frac{r_{0}}{R_{e}}\right) ^{2}+\ldots\ ,  \label{expan1}
\end{equation}
where $\lambda_{\infty}\equiv\mu_{\infty }$ is the equilibrium
chemical potential for the infinite nuclear matter. We will apply
the Gibbs -- Duhem relation
\begin{equation}
dP_{\mathcal{V}}=\varrho_{\mathcal{V}}\,d\lambda\ .  \label{gd}
\end{equation}
Using the generalized Laplace equation (\ref{genlap}) and
Eqs.~(\ref{sigmaeq}) and (\ref{expan1}), we rewrite Eq.~(\ref{gd}) as 
\[
d\left(\frac{2\sigma_{\infty }}{R_{e}}-\frac{2\sigma_{\infty}\xi}
{R_{e}^{2}}+\ldots\right)\! =\!
\left( \rho_{\infty}+\rho_{1}\frac{r_{0}}{R_{e}}+
\rho_{2}\frac{r_{0}^{2}}{R_{e}^{2}}+\ldots\right) 
\]
\begin{equation}
\times\, d\left(\lambda_{\infty}+\lambda_{1}\frac{r_{0}}{R_{e}}+
\lambda_{2}\frac{r_{0}^{2}}{R_{e}^{2}}+\ldots\right)\ .   \label{gd1}
\end{equation}
Nuclear incompressibility $K_{\infty }$ in terms of expansion (\ref{expan1})
reads
\begin{equation}
K_{\infty}=9\left.\frac{\partial P_{\mathcal{V}}}{\partial\varrho_{\mathcal{V}}}
\right\vert_{\varrho _{\mathcal{V}}=\rho_{\infty}}\!=
9\left.\varrho_{\mathcal{V}}\frac{\partial\lambda}{\partial\varrho_{\mathcal{V}}}
\right\vert_{\varrho_{\mathcal{V}}=\rho_{\infty}}\!=
9\rho_{\infty}\frac{\lambda_{1}}{\rho _{1}}\ .  \label{k}
\end{equation}%
Equating in (\ref{gd1}) the terms of the same order in curvature $R_{e}^{-1}$
and taking the incompressibility definition from Eq.~(\ref{k}), one
obtains the following relations
\begin{equation}
\rho_{1}=18\frac{\sigma_{\infty }}{K_{\infty}\,r_{0}}\ ,
\ \ \lambda_{1}=2\frac{\sigma_{\infty}}{\rho_{\infty}\,r_{0}}  \label{rel1}
\end{equation}
and
\begin{equation}
\xi =-\,9\frac{\sigma_{\infty}}{K_{\infty}\,\rho_{\infty}}-
\frac{\lambda_{2}}{\lambda _{1}}r_{0}\ .  \label{rel2}
\end{equation}
Equation (\ref{rel2}) gives an idea how the Tolman length $\xi$ depends on
the incompressibility $K_{\infty}$ and the surface tension coefficient $\sigma$.
In particular, if the second order $\sim R_{e}^{-2}$ correction in the chemical
potential $\lambda$ of Eq.~(\ref{expan1}) is negligible, namely,
\[
\lambda =\lambda_{\infty }+\lambda_{1}\frac{r_{0}}{R_{e}},
\]
we obtain from Eq.~(\ref{rel2}) the following important relation
\begin{equation}
\xi \approx -\,9\,\frac{\sigma_{\infty}}{K_{\infty}\,\rho _{\infty}}\ .
\label{tolman1}
\end{equation}
That means that the Tolman length disappears in the case of incompressible
Fermi liquid with $K_{\infty}\rightarrow\infty$. We note also the relation of the
surface tension coefficient $\sigma$ to the incompressibility $K_{\infty}$
and the diffuseness parameter $a$ of the nuclear surface layer \cite{cast80}
\begin{equation}
\sigma_{\infty}\approx\frac{1}{18}K_{\infty}\,\rho_{\infty}a\ .
\label{sigma2}
\end{equation}%
Comparing Eqs.~(\ref{tolman1}) and (\ref{sigma2}) we conclude that
\[
\xi\approx -\,a/2\ .
\]
This result leads to the conclusions that the nuclear Tolman length is
negative and the non-zero value of $\xi$ requires the finite diffuse layer.

\section{Conclusions}

Considering a small two-component, charged droplet with a finite diffuse
layer, we have introduced a formal dividing surface of radius $R$ which
splits the droplet onto volume and surface parts. The corresponding
splitting was also done for the binding energy $E$. Assuming that the
dividing surface is located close to the interface, we are then able to
derive the surface energy $E_{\mathcal{S}}$. In general, the surface energy
$E_{\mathcal{S}}$ includes the contributions from the surface tension $\sigma$
and from the binding energy of $A_{\mathcal{S}}$ particles located within
the surface layer. The equimolar surface and thereby the actual physical
size of the droplet are derived by the condition
$\varrho _{\mathcal{S}}\lambda +\varrho_{-,\mathcal{S}}\lambda _{-}=0$
which means that the latter contribution is excluded from the surface energy providing
$E_{\mathcal{S}}\propto\sigma$.

In a small nucleus, the diffuse layer and the curved interface affect the
surface properties significantly. In agreement with Gibbs-Tolman concept 
\cite{tolm49,gibbs}, two different radii have to be introduced in this case.
The first radius, $R_{s}$, is the surface tension radius (Laplace radius)
which provides the minimum of the surface tension coefficient $\sigma$ and
the fulfillment of the Laplace relation (\ref{p3}) for capillary pressure.
The another one, $R_{e}$, is the equimolar radius which corresponds to the
equimolar dividing surface due to the condition (\ref{emolar}) and defines
the physical size of the sharp surface droplet, i.e., the surface at which
the surface tension is applied. The difference of two radii $R_{e}-R_{s}$ in
an asymptotic limit of large system $A\rightarrow\infty$ derives the
Tolman length $\xi$. That means the presence of curved surface is not
sufficient for the presence of the curvature correction in the surface
tension. The finite diffuse layer in the particle distribution is also
required. We point out that the Gibbs-Tolman theory allows to treat a liquid
drop within thermodynamics with minimum assumptions. Once the binding energy
and chemical potential of the nucleus are known its equimolar radius, radius
of tension and surface energy can be evaluated using the equation of state
for the infinite nuclear matter. For a symmetric liquid the
value of Tolman length is about of half of the diffuseness parameter $a$ for
the nuclear surface layer. We have also established the relation of the
macroscopic energy coefficients in the liquid drop model expansion
Eq.~(\ref{eldm1}) to the nuclear matter parameters.

The sign and the magnitude of the Tolman length $\xi$ depend on the
interparticle interaction. We have shown that the Tolman length is negative
for a nuclear Fermi liquid drop. As a consequence, the curvature correction
to the surface tension leads to the hindrance of the yield of light
fragments at the nuclear multifragmentation in heavy ion collisions. We have
also shown that the Tolman length is sensitive to the neutron excess and its
absolute value growth significantly with growing asymmetry parameter $X$.

\appendix
\section{Relation of nuclear matter EOS to the characteristics of finite
nuclei}

We will start from the nuclear matter EOS given by Eq. (\ref{fmatter}) and
take into consideration the relations (\ref{mKJ}) and (\ref{b0}) and the
following higher order coefficients 
\[
K_{3}=6K_{\infty}+27\left.\rho^{3}\frac{\partial^{3}E_{\infty}/A}
{\partial\rho^{3}}\right\vert_{\rho=\rho_{\infty},\,x=0}\ ,
\]
\begin{equation}
L_{\infty}=\frac{3}{2}\left.\rho\frac{\partial^{3}E_{\infty}/A}
{\partial\rho\partial x^{2}}\right\vert_{\rho=\rho_{\infty},\,x=0}\ ,
\label{homom}
\end{equation}
\begin{equation}
K_{\mathrm{sym}}=\frac{9}{2}\left.\rho^{2}\frac{\partial^{4}E_{\infty}/A}
{\partial\rho^{2}\partial x^{2}}\right\vert_{\rho=\rho_{\infty},\,x=0}\ ,
\label{ksym1}
\end{equation}
for the expansion (\ref{fmatter}). Here $K_{3}$ is the bulk anharmonicity
coefficient, $L_{\infty}$ is the density-symmetry coefficient (symmetry
energy slope parameter), $K_{\mathrm{sym}}$ is the symmetry energy curvature
parameter. Using (\ref{sigmaeq}), we write also 
\[
\sigma\approx\sigma_{\infty}\left( 1-2\xi/R_{e}\right)\ ,
\]
\begin{equation}
\sigma_{\infty}\approx\sigma _{0}+\sigma_{-}X^{2}\ ,
\ \ \ \xi \approx\xi_{0}+\xi_{-}X^{2}  \label{sigma1x}
\end{equation}
and 
\begin{equation}
a_{V}=\mu _{\infty}\ ,
\ \ \ b_{V}=b_{\infty}\ .  \label{av1x}
\end{equation}

Using the conditions (\ref{eosmatt}) for the chemical potentials and both
relations (\ref{lambda01}) and (\ref{mu01p}), we obtain
\begin{widetext}
\[
\frac{\rho-\rho_{\infty}}{\rho_{\infty}}\approx
A^{-1/3}\frac{6a_{S}}{K_{\infty}}+X^{2}\left[ -\frac{3L_{\infty}}{K_{\infty }}+
A^{-1/3}\left\{ \frac{6(b_{S}-2a_{S}L_{\infty}/K_{\infty})}{K_{\infty}}
\left( 1-\frac{L_{\infty}}{b_{\infty }}\right)-\frac{6a_{S}}{K_{\infty}^{2}}
\left[ L_{\infty}\left( 1-\frac{K_{3}}{K_{\infty}}\right)+
K_{\mathrm{sym}}\right]\right\}\right]
\] 
and 
\begin{equation}
a_{S}=4\pi r_{0}^{2}\sigma_{0}\ ,
\ \ \ b_{S}=4\pi r_{0}^{2}\left(\sigma_{-}+
\frac{2L_{\infty}}{K_{\infty}}\,\sigma_{0}\right)\ ,  
\ \ \ a_{c}=-8\pi r_{0}\sigma_{0}\left(\xi_{0}+
\frac{3\,\sigma_{0}}{K_{\infty}\rho_{\infty}}\right)\ ,
\label{as1x}
\end{equation}
\begin{equation}
b_{c}=-8\pi r_{0}\sigma_{0}\left\{\xi_{-}+\left(\frac{L_{\infty}}{K_{\infty}}+
\frac{\sigma_{-}}{\sigma_{0}}\right)\xi_{0}+
\frac{3\,\sigma_{0}}{K_{\infty}\rho_{\infty}}\left[\frac{L_{\infty}}{K_{\infty}}
\left( 4+\frac{K_{3}}{K_{\infty}}\right)-
\frac{K_{\mathrm{sym}}}{K_{\infty}}\right]+
\frac{3\,\sigma_{-}}{K_{\infty}\rho_{\infty}}\left( 2+
\frac{K_{\infty}\sigma_{-}}{2b_{\infty}\sigma_{0}}\right)\right\}\ .
\label{bc1x}
\end{equation}
Here we have assumed $A^{-1/3}\ll 1$. The equimolar, $R_{e}$, and Laplace, $R_{s}$,
radii defined by Eqs.~(\ref{evol}) and (\ref{p3}) read
\[
R_{e}\approx r_{0}A^{1/3}\left[ 1-
A^{-1/3}\frac{8\pi r_{0}^{2}\sigma_{0}}{K_{\infty}}
\right. 
\]
\begin{equation}
\left. 
+\ X^{2}\left[\frac{L_{\infty}}{K_{\infty}}-
A^{-1/3}\left\{\frac{8\pi r_{0}^{2}\sigma_{-}}{K_{\infty}}\left( 1-
\frac{L_{\infty}}{b_{\infty}}+\frac{K_{\infty }}{3\,\mu_{\infty}}\right)+
\frac{8\pi r_{0}^{2}\sigma_{0}}{K_{\infty}}\left[\frac{L_{\infty}}{K_{\infty}}\left( 3+
\frac{K_{3}}{K_{\infty}}\right) -\frac{K_{\mathrm{sym}}}{K_{\infty}}
\right]\right\}\right]\right]\ ,  \label{re1x}
\end{equation}
\[
R_{s}\approx r_{0}A^{1/3}\left[ 1-
A^{-1/3}\left(\frac{\xi_{0}}{r_{0}}+
\frac{8\pi r_{0}^{2}\sigma_{0}}{K_{\infty}}\right)
\right. 
\]
\begin{equation}
\left.
+\ X^{2}\left[\frac{L_{\infty}}{K_{\infty}}-
A^{-1/3}\left\{\frac{\xi_{-}}{r_{0}}+\frac{8\pi r_{0}^{2}\sigma_{-}}{K_{\infty}}\left( 1+
\frac{K_{\infty}}{2b_{\infty}}\frac{\sigma_{-}}{\sigma_{0}}\right)+
\frac{8\pi r_{0}^{2}\sigma_{0}}{K_{\infty}}\left[\frac{L_{\infty}}{K_{\infty}}\left( 3+
\frac{K_{3}}{K_{\infty}}\right)-\frac{K_{\mathrm{sym}}}{K_{\infty}}
\right]\right\}\right]\right]\ .  \label{rs1x}
\end{equation}
Using the derivations of $R_{e}$ and $R_{s}$, one obtains
\begin{equation}
R_{e}-R_{s}\approx\xi_{0}+\left[
\xi_{-}+\frac{3\sigma_{-}}{b_{\infty}\rho_{\infty}}\left(
\frac{\sigma_{-}}{\sigma_{0}}+\frac{2L_{\infty}}{K_{\infty}}-
\frac{2b_{\infty}}{3\mu_{\infty}}\right)\right] X^{2}=
\xi +\left[\frac{3\sigma_{-}}{b_{\infty}\rho_{\infty}}\left(
\frac{\sigma_{-}}{\sigma_{0}}+\frac{2L_{\infty}}{K_{\infty}}-
\frac{2b_{\infty}}{3\mu_{\infty}}\right)\right] X^{2}\ .
\label{R_e_s}
\end{equation}

To describe separately the neutron and proton density distributions we
introduce the neutron radius, $R_{n}$, and the proton radius, $R_{p}$, as
the dividing radii with zero value for the corresponding surface densities
$\varrho_{n,\mathcal{S}}=(\varrho_{\mathcal{S}}+\varrho_{-,\mathcal{S}})/2$ and
$\varrho_{p,\mathcal{S}}=(\varrho_{\mathcal{S}}-\varrho_{-,\mathcal{S}})/2$ :
\[
\left.\varrho_{n,\mathcal{S}}\right\vert_{R=R_{n}}=0\ ,\ \ \ \ \ \ 
\left.\varrho_{p,\mathcal{S}}\right\vert_{R=R_{p}}=0\ .
\]
The value of neutron skin $r_{np}=R_{n}-R_{p}$ is then written as
\begin{equation}
r_{np}=R_{n}-R_{p}\approx X\left[ -\frac{2\,\sigma_{-}}{b_{\infty}\rho_{\infty}}+
A^{-1/3}\left\{ 4\pi r_{0}^{2}\sigma_{0}\frac{4}{3b_{\infty}}\left(
\xi_{-}+\xi_{0}\frac{\sigma_{-}}{\sigma_{0}}\right) +4\pi r_{0}^{2}\sigma_{-}\left[
\frac{2\,\sigma_{-}}{b_{\infty}^{2}\rho_{\infty}}+\frac{4\,\sigma_{0}}
{b_{\infty}^{2}\rho_{\infty}}\left(\frac{L_{\infty}}{K_{\infty}}+
\frac{3b_{\infty}}{K_{\infty}}\right)\right]\right\}\right]\ .   \label{rnp1x}
\end{equation}
\end{widetext}

\end{document}